\documentclass[showpacs,amsmath,amssymb,twocolumn,prl,superscriptaddress,notitlepage,floatfix]{revtex4-2}

\usepackage{graphics}
\usepackage{amsmath,amssymb,amsthm,mathrsfs,amsfonts,dsfont}
\usepackage{subfigure, epsfig}
\usepackage{braket}
\usepackage{bm}
\usepackage{enumerate}
\usepackage{physics}
\usepackage{diagbox}

\usepackage{booktabs}
\usepackage[linesnumbered, boxed,commentsnumbered, ruled]{algorithm2e}
\usepackage{makecell}      
\usepackage{bbding}        

\usepackage[normalem]{ulem}

\usepackage{pgfplots}
\pgfplotsset{width=7cm,compat=1.9}

\usepgfplotslibrary{external}

\usepackage{algpseudocode}
\usepackage{color}
\usepackage{comment}

\usepackage[colorlinks=true, linkcolor=blue]{hyperref}
\graphicspath{{./figure/}}

\newcommand{\mc}{\mathcal}
\newcommand{\mb}{\mathbf}

\newcommand{\id}{\mathbb{I}}
\newcommand{\sw}{\mathbb{S}}

\newcommand{\pr}{\mathrm{Pr}}

\newcommand{\comments}[1]{}

\begin{document}

\title{Directly Estimating Mixed-State Entanglement with Bell Measurement Assistance}

\author{Gong-Chu Li}
\affiliation{CAS Key Laboratory of Quantum Information, University of Science and Technology of China, Hefei, Anhui 230026, China.}
\affiliation{CAS Center For Excellence in Quantum Information and Quantum Physics, University of Science and Technology of China, Hefei, Anhui 230026, China.}
\affiliation{Hefei National Laboratory, Hefei 230088, China}
\author{Lei Chen}
\author{Si-Qi Zhang}
\author{Xu-Song Hong}
\affiliation{CAS Key Laboratory of Quantum Information, University of Science and Technology of China, Hefei, Anhui 230026, China.}
\affiliation{CAS Center For Excellence in Quantum Information and Quantum Physics, University of Science and Technology of China, Hefei, Anhui 230026, China.}
\author{You Zhou}
\email{you\_zhou@fudan.edu.cn}
\affiliation{Key Laboratory for Information Science of Electromagnetic Waves (Ministry of Education), Fudan University, Shanghai 200433, China}
\affiliation{Hefei National Laboratory, Hefei 230088, China}
\author{Geng Chen}
\email{chengeng@ustc.edu.cn}
\author{Chuan-Feng Li}
\email{cfli@ustc.edu.cn}
\author{Guang-Can Guo}
\affiliation{CAS Key Laboratory of Quantum Information, University of Science and Technology of China, Hefei, Anhui 230026, China.}
\affiliation{CAS Center For Excellence in Quantum Information and Quantum Physics, University of Science and Technology of China, Hefei, Anhui 230026, China.}
\affiliation{Hefei National Laboratory, Hefei 230088, China}

\begin{abstract}

\comments{
 Entanglement plays a fundamental role in quantum physics and information processing. 
 Here we directly estimate the mixed-state entanglement with random unitary evolution in a photonic system. As a supplement to the traditional projective measurements, we add the Bell measurement on qubit-pair to enrich the previous randomized measurement scheme,
 which is \textit{no-go} in this task with only local unitary evolution. The scheme can be scaled up to $n$-qubit via only
the Bell measurement on qubit-pairs. 
The estimator can be directly read from a few consecutive outcomes, and at the same time, its performance is more robust to system error and noise compared to other schemes based on shadow estimation. 
Our protocol and demonstration advance the direct characterization of quantum states in practice.
}

Entanglement plays a fundamental role in quantum physics and information processing. Here, we develop an unbiased estimator for mixed-state entanglement in the few-shot scenario and directly estimate it using random unitary evolution in a photonic system. As a supplement to traditional projective measurements, we incorporate Bell measurements on qubit-pairs, enriching the previous randomized measurement scheme, which is \textit{no-go} in this task with only local unitary evolution. The scheme is scalable to $n$-qubits via Bell measurements on qubit-pairs. 
The estimator can be derived directly from a few consecutive outcomes while exhibiting greater robustness to system errors and noise compared to schemes based on shadow estimation. And we find that, under a fixed measurement resource, the way with more versatile measurement settings with fewer repeats per setting is more efficient. Our protocol and demonstration advance the direct characterization of quantum states in practice.

\end{abstract}

\maketitle


\textit{Introduction---}
Characterization and quantification of entanglement is one of the core topics in the field and is challenging in general \cite{Horodecki2009entanglement,GUHNE2009detection,Friis2019Reviews}. 
Methods like entanglement witness \cite{chruscinski2014entanglement, liu2022fundamental} need the prior knowledge of the state form, and full tomography not only needs extensive resources but also may become biased for the later calculation of the entanglement \cite{silva2017investigating}. Moreover, both of them are not suited for the few-shot scenario.
One widely employed entanglement measure without the prior knowledge is the negativity $\mathcal{N}= \sum_{\lambda <0}|\lambda|$ \cite{Vidal2002measure,Plenio2005Negativity}, with $\lambda$ the eigenvalues of $\rho^{T_B}_{AB}$, which is by the partial transpose (PT) operation $T_B$ on the joint density matrix. Such entanglement measures are highly nonlinear and complex \cite{Horodecki2009entanglement,bengtsson2017geometry}, and thus cannot be directly evaluated.
Thus, here we aim to build up an unbiased estimator to quantify entanglement negativity, which can be directly read-out from the collected data, even in a few-shot scenario.

\comments{
\begin{figure}[]
    \centering
    \includegraphics[width=0.5\textwidth]{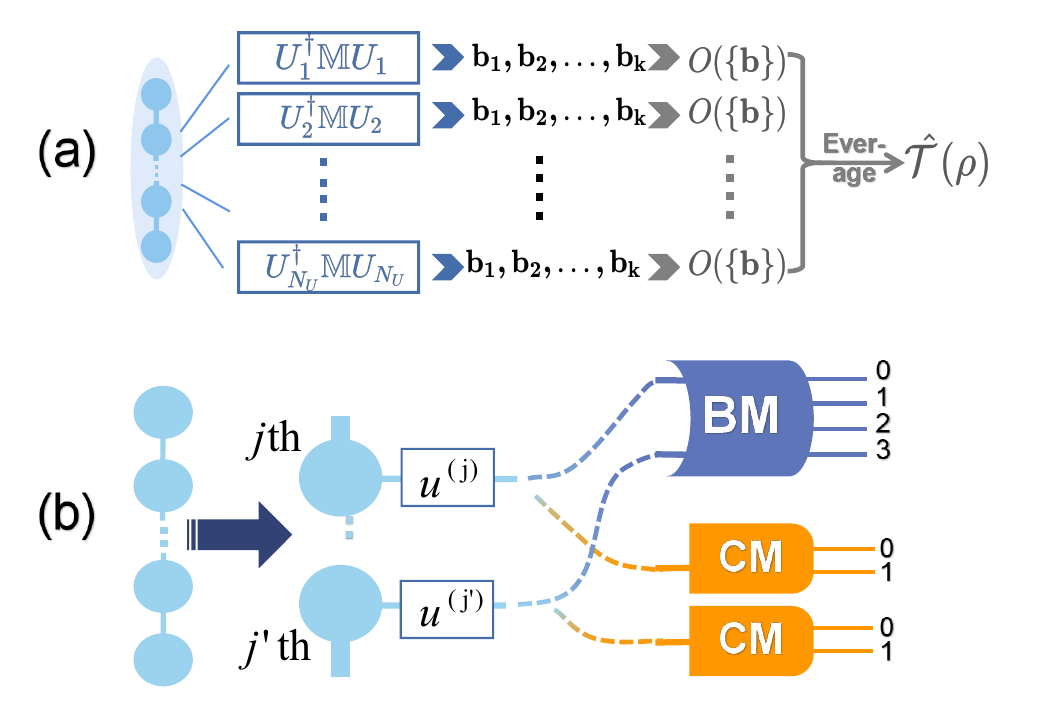}
    \caption[]{
    \textit{(a) General procedure of CS, RM, and FSRM:} All schemes subject, say, an $n$-qubit state to random unitary rotation before measurement (here $U=\bigotimes_j u^{(j)}$), resulting in an $n$-bit string $\mb{b}$ as the outcome. The CS scheme requires both $\mb{b}$ and $U$ information, while the RM scheme demands multiple repetitions of the same unitary to estimate conditional probabilities across all bases, i.e.,  $\{\mathrm{Pr}(\mb{b}|U)\}$. In contrast, the FSRM scheme relies solely on $k$-shot outcomes $\mb{b}_1,\dots,\mb{b}_k$ per unitary when predicting some $k$-th order properties, without the need for detailed unitary information.
    \textit{(b) Demonstration of the BM-enhanced RM:} We slice the state into pairs of qubits. After the random unitary rotation shown in (a), all pairs undergo a Bell-basis measurement (BM) or a computational-basis measurement (CM) with some probability. For BM, we record bases with labels ${0, 1, 2, 3}$, and for CM, we record bases with labels ${00, 01, 10, 11}$ for post-processing. The overall measurement basis can be represented by an $n$-bit string $\vec{s}$.
    }
    \label{Fig_demo}
\end{figure}
}

\begin{figure}[]
    \centering
    \includegraphics[width=0.5\textwidth]{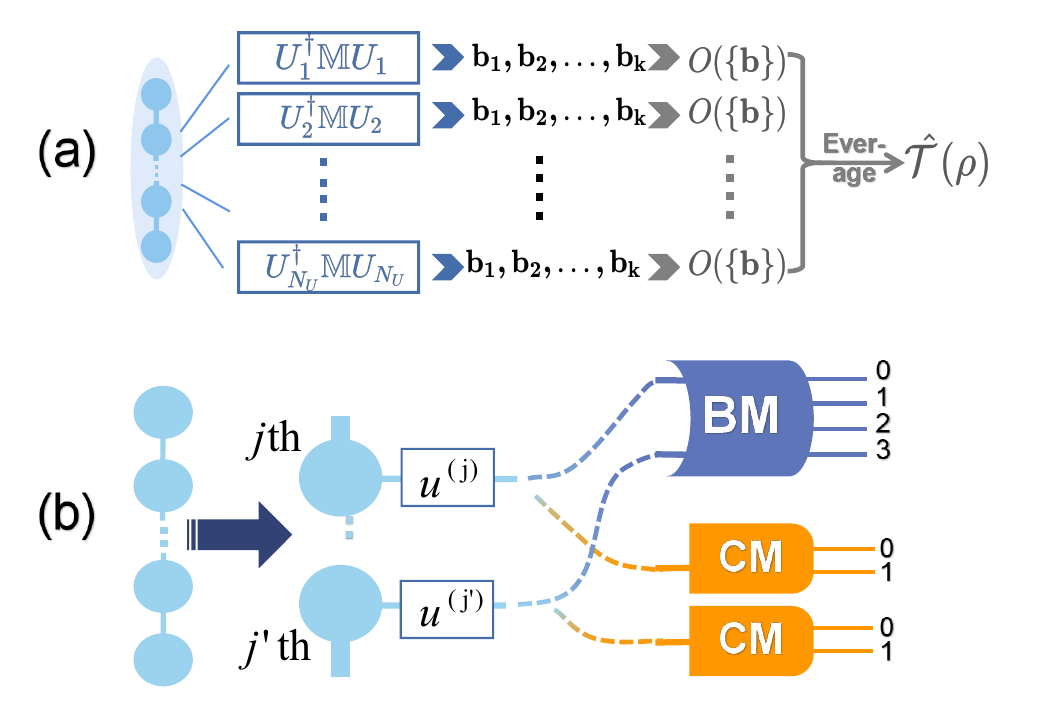}
    \caption{
     \textit{(a) Demonstration of Few-shot Randomized Measurement (FSRM).}
     Few-shot randomized measurement involves performing $N_U$ measurement settings, $U_1^\dagger\mathbb{M} U_1$ to $U_{N_U}^\dagger \mathbb{M} U_{N_U}$, where $\mathbb{M}$ is a fixed measurement and the unitary is randomly chosen. The \textit{few-shot} aspect means collecting only $k$ shots for each setting, with $k=3$ for estimating mixed-state entanglement. If $\mathbb{M}$ has $D$ outcomes, each shot is a number from 0 to $D-1$. These $k$ shots are then used to obtain an unbiased estimate of the desired property, $O(\{\mathbf{b_i}\})$.
     \textit{(b) Demonstration of the BM-enhanced FSRM:} We slice the state into pairs of qubits. After the random unitary rotation shown in (a), all pairs undergo a Bell-basis measurement (BM) or a computational-basis measurement (CM) with some probability, i.e., $\mathbb{M}\in \{\mathrm{CM, BM}\}$ . For BM, we record bases with labels ${0, 1, 2, 3}$, and for CM, we record bases with labels ${00, 01, 10, 11}$ for post-processing. The overall measurement basis can be represented by an $n$-bit string $\vec{s}$.
     These recorded bases as the measurement shots get an unbiased estimation with the process in (a)
    }
\label{Fig_demo}
\end{figure}

Recently, measurement schemes based on random unitary (RU) evolution \cite{elben2023randomized} 
, especially shadow estimation \cite{aaronson2019shadow,huang2020predicting} suiting for few-shot scenario, and develop very fast, with proposals emerging to directly estimate a few nonlinear functions in the form of $\mathcal{T}_k(\rho)=\Tr(T_k\rho^{\otimes k})$ for a $k$-th order function in a sequential manner, without simultaneous control on many copies \cite{van2012Measuring,Elben2018Random,Brydges2019Probing,elben2020mixedstate,singlezhou}. In the context of entanglement quantification, a low-order approximation of the negativity \cite{neven2021symmetry,Yu2021Optimal}, $p_3$-PPT, was proposed, represented by $\mathcal{N}_3=p^2_2-p_3$ \cite{elben2020mixedstate}, where $p_k=\Tr[(\rho^{T_B}_{AB})^{k}]$ denotes the $k$-th moment of the PT density matrix \cite{Bose2018Machine,singlezhou}. Importantly, such moments can be expressed as a $k$-order function $T_k$ involving permutation operators \cite{Bose2018Machine,singlezhou,elben2020mixedstate}, allowing the RU scheme to be directly applied for their estimation.

In the framework of shadow estimation, the classical shadow (CS) of a quantum state can be constructed from RU \cite{huang2020predicting}, and the estimation of PT-moment $p_k$ has been demonstrated \cite{elben2020mixedstate,zhang2021experimental}. However, CS scheme generally requires the precise knowledge of the random unitaries used in the estimation process, as reflected by the post-processing strategy. 
This limitation could lead to a mismatch and estimation error in real experiments.


On the other hand, there is an alternative RU scheme \cite{van2012Measuring,Elben2019toolbox} and here for distinction we denote it as randomized measurements (RM), which mainly differ from CS by introducing multi-shot \cite{Helsen2023Thrifty,zhou2023performance} in one round with a different classical post-processing strategy \cite{Elben2019toolbox}. 
It is worth noting that while RM has shown promises in estimating various properties \cite{Elben2020Cross,Elben2020topological,Cian2020Chern,rath2021Fisher,yu2021Fisher,Zhenhuan2022correlation}, it has not yet been applied to 
mixed-state entanglement. 
The main obstacle is a \textit{no-go} theorem by You Zhou et.~al.~claiming that the 3-moment $p_3$ can not be estimated by local random unitary in the form of $U_A\otimes U_B$ and computational-basis measurement(CM) \cite{singlezhou}. 
Additionally, RM methods are not developed for the few-shot scenario, limiting their efficiency. 

Here, in this work, we experimentally demonstrate the RM scheme enhanced by bi-qubits Bell measurement (BM) is able to detect mixed-state entanglement by estimating $p_2$ and $p_3$ of photonic states. To make the post-processing of the RM scheme more efficient, we adapt it to the few-shot scenario, referred to as the few-shot randomized measurement (FSRM).
We find that compared to the CS scheme, the RM scheme is more robust to noise in practice. 
In addition, we discuss the improvement of sampling efficiency with more versatile measurement settings and fewer repeats per setting under the constraints of quantum resources.
Our work advances the RM scheme in the characterization of photonic quantum states, especially in the context of entanglement quantification, and could be extended to multipartite systems in principle. 

\textit{BM-ehanced FSRM scheme---}
Let us first recast the basics of randomized measurement (RM), and here we mainly focus on an $n$-qubit system. 
In each round, the RM scheme samples a random unitary from some ensemble, 
e.g. local random ones $U=\otimes_j u^{(j)}$ \cite{Elben2019toolbox}. One samples $U$ for $N_U$ rounds. In each round, one repeats the evolution of 
the chosen $U$ 
for $N_M$ shots.  
One post-processes the multiplication of conditional probabilities $\mathrm{Pr}(\mb{b}|U)$ to estimate some nonlinear functions like the purity \cite{Brydges2019Probing}, 
\begin{equation}\label{eq:RMmean}  
\begin{aligned}
\mathcal{T}_k(\rho) = \mathbb{E}_U \sum_{\{\mb{b}_i\}} O_k(\{\mb{b}_i\})\ \prod_{i=1}^k\mathrm{Pr}(\mb{b}_i|U).
\end{aligned}
\end{equation}

Here $O_k$ serves as a proper post-processing function that projects $k$-shot result into a real number, with each shot as a $n$-bit string. 
In practice, to obtain a reasonable accurate $\pr(\mb{b}|U)$, a large number of shots, typically $N_M \sim 10^2$ for a single-qubit experiment, are required, which seems to hinder RM in the few-shot scenario. 

Actually, inspired by the CS scheme, one can construct an unbiased estimator of a $k$-th order function $\mc{T}_k$ within the RM framework using very few shots \cite{singlezhou,Zhenhuan2022correlation}. In particular, the shot number can be taken to be $N_M=k$, i.e., its minimal value. By denoting in total $k$ successive shots in the experiment as $\{\mb{b}_i\}$.
The unbiased estimator reads $\hat{\mc{T}_k}=O_k(\{\mb{b}_i\})$ with 
\begin{equation}
    \mc{T}_k(\rho) = \mathbb{E}_U\mathbb{E}_{\{\mb{b}_i\}}O_k(\{\mb{b}_i\}),
    \label{Eq_FSRM}
\end{equation}
which is guaranteed by Eq.~\eqref{eq:RMmean}. We denote such modification of RM about the post-processing strategy as few-shot randomized measurements (FSRM).The process for the FSRM see Fig.\ref{Fig_demo}(a).
This estimator is then averaged for $N_U$ times to reduce the statistical variance.
Note that, unlike the estimation of linear observables, we give an unbiased estimation of $\mc{T}_k(\rho)$ from every $k$ shots.

It is worthy to mention that FSRM enjoys features and advantages like robustness and efficient post-processing from RM, and few-shot from CS. A detailed comparison among CS, RM and FSRM schemes is left in Sec.I of Ref.~\cite{supp}.

In the context of PT-moment estimation, 
one needs $p_2$ and $p_3$ to estimate mixed-state entanglement.
For the $2$-th order, it can be expressed as $p_2=\tr(\sw \rho_{AB}^{\otimes 2})$, with $\sw$ being the swap operator on 2 copies. 
Suppose one applies a local random unitary ensemble in RM, the post-processing function can be taken as $O_2(\mb{b}_1,\mb{b}_2) = d(-2)^{h(\mb{b}_1,\mb{b}_2)}$ with $h(\mb{b}_1,\mb{b}_2)$ being the Hamming distance between two n-bit strings \cite{Elben2018Random, Elben2019toolbox}. 

For the 3-order moment, 
We follow the post-processing given in Ref.~\cite{singlezhou}, which in general needs global random unitary, or BM between subsystem $A$ and $B$ after local random unitary.
As illustrated in Fig.~\ref{Fig_demo}(b), for an n-qubit state, BM on qubit-pairs is sufficient.
We denote the outcome of BM on $(j,j')$ qubit-pair as $\beta^{(j,j')}=\{0,1,2,3\}$.
The specific partition is labeled by an n-bit string $\vec{a}$, with $a_i=1(0)$ for qubit-i in $A(B)$, respectively. Additionally, we use another n-bit string $\vec{s}$ to indicate whether the qubit is involved in BM or not.  Of course, $\vec{s}$ should have even 
number of `1's, and we randomly sample in total $2^{n-1}$ possible $\vec{s}$ configurations with equal probability. 
In short, the corresponding estimator of $p_3$ is $O_3(\mb{b}_1,\mb{b}_2,\mb{b}_3) = \sum_{\vec{s}}\pr(\vec{s})O_3(\mb{b}_1,\mb{b}_2,\mb{b}_3|\vec{s})$ with $\pr(\vec{s})=1/2^{n-1}$ and 

\begin{equation}\small
\begin{aligned}
O_3(\mb{b}_1,\mb{b}_2,\mb{b}_3|\vec{s}) = \frac1{2}(-1)^{\vec{a}\cdot\vec{s}}& \prod_{j:s_j=1} f(j,j')
    \prod_{j:s_{j}=0} g(j).
\end{aligned}
    \label{p3_RM}
\end{equation}
Here $f$ and $g$ functions are for the BM and CM outcome of qubit-pair $(j,j')$ and qubit $j$, determined by the measurement configuration vector $\vec{s}$. And they show respectively
\begin{equation}\label{eq:f}
\begin{aligned}
    &f(j,j')=1-(-2)^{\mathrm{wt}(\beta_1^{(j,j')},\beta_2^{(j,j')},\beta_3^{(j,j')})}\\
    \text{and } &g(j)=1+(-2)^{[\mathrm{wt}(b_1^{(j)},b_2^{(j)},b_3^{(j)})-1]},
\end{aligned}
\end{equation}
where $\mathrm{wt}()$ counts the number of same elements for $\beta$ and $b$.
More details are left in Sec.II of Ref.~\cite{supp}.  

In the case of 2-qubit, with $A$ and $B$ containing each of them, i.e., $\vec{a}=10$. The vector $\vec{s}=\{00,11\}$, which indicates that one conducts CM and BM with equal probability on 2-qubit. By Eq.~\eqref{p3_RM}, the post-processing is just 
\begin{equation}\label{eq:f2q}
O_3=\left\{
\begin{aligned}
    \frac1{2}&g(1)g(2),\ &\vec{s}=00,\\
    -\frac1{2}&f(1,2),\ &\vec{s}=11.
\end{aligned}
\right.
\end{equation}

In summary, the procedure for estimating $p_2$ and $p_3$ is as follows. We implement random local unitaries $U=\otimes_j u^{(j)}$ on the state $\rho$ and randomly select a measurement setting $\vec{s}$. For each sampled unitary $U$ and $\vec{s}$, we repeat the process three times to collect measurement outcomes $\{\mathbf{b}_{i=1}^3\}$, and finally calculate the function $O_3(\mathbf{b}_1,\mathbf{b}_2,\mathbf{b}_3|\vec{s})$ in Eq.~\eqref{p3_RM} as the estimation of $p_3$. For $p_2$, we only need the measurement setting all being CM, say $\vec{s}=\vec{0}$, and two outcomes are sufficient to obtain $O_2(\mathbf{b}_1,\mathbf{b}_2)$ as an estimation.
By repeating this process many times, we average the results to obtain the final estimations of $p_2$ and $p_3$. The post-processing here is straightforward and time-saving compared to the original RM scheme and CS scheme, which involves calculating multiple shadow snapshots and their multiplications.

\begin{figure}[htbp]
    \centering
    \includegraphics[width=0.4\textwidth]{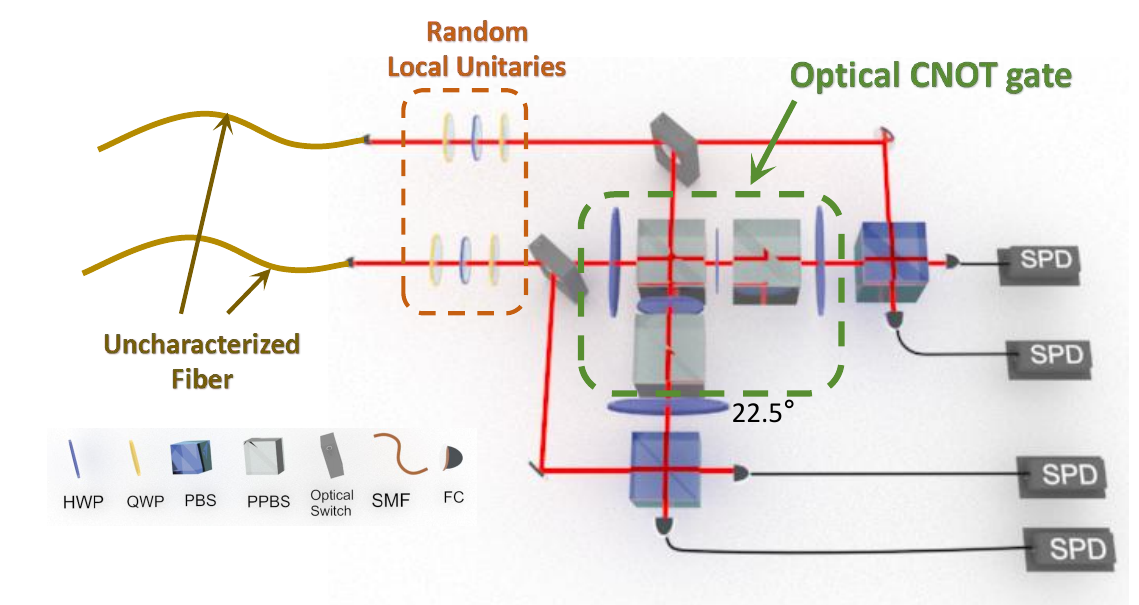}
    \caption{\textit{Sketch of the measurement setup.}
    Entangled biphoton states are transmitted to the measurement setup through uncharacterized fibers. Each photon within the setup undergoes a randomized unitary, selected from a Circular Unitary Ensemble (CUE) implemented using a set of quarter-half-quarter waveplates (depicted by the orange circle). Two optical switches, each featuring one input and two outputs, determine whether the pair of photons will be measured in computational basis (CM) along the outer path or in Bell basis (BM) along the inner path, with a probability of 1/2. Along the outer path, the CM setup consists of two polarizing beam splitters (PBSs). In contrast, the inner path includes an all-optical CNOT gate (depicted by the green circle). This setup, combined with an additional Hadamard gate (22.5° half-wave-plate) and CM, forms a complete BM. Finally, single-photon detectors (SPDs) register the pair of photons. Further details can be found in Sec.III of Ref.~\cite{supp}.
    }
    \label{Fig_setup}
\end{figure}
\textit{Experimental Results---}
In our experiments, states with polarization entanglement are produced through the parameter-down-conversion (PDC) process in the periodically poled potassium titanyl phosphate (ppKTP) crystal within a Sagnac interferometer \cite{kim2006phase}. Pure or mixed states could be easily produced by adjusting the pump laser. In the experiments, we produced both a mixed state and a maximally entangled pure state, as detailed in Sec.IV of Ref.~\cite{supp}. The produced photons are then sent into two uncharacterized single-mode fibers (SMF) before measurement.

The measurement setup is illustrated in Fig.~\ref{Fig_setup}, where two photons undergo local random unitaries using Quarter-Half-Quarter (Q-H-Q) waveplates \cite{englert2001universal}. 
Because BM could be decomposed with a CNOT gate and a 
Hadamard gate before CM, we can streamline the process of random CM and BM by randomly inserting such gates between qubit-pairs and finally conducting CM.
In the experiment, two optical switches, each with one input and two outputs made by motorized mirrors control the ratio of photon pairs that go through the inner or outer path, ensuring a 50\%-50\% ratio between CM (inner path) and BM (outer path). CM involves measuring horizontal and vertical polarizations in our experiments, resulting in a string of $\{0,1\}^n$. On the inner path, an additional all-optical CNOT gate 
consisting of three partially-polarized-beam-splitters (PPBSs)  \cite{o2003demonstration, kiesel2005linear, okamoto2005demonstration} and a Hadamard gate on the control part were implemented.

\begin{figure}[htbp]
    \raggedleft
    \includegraphics[width=0.47\textwidth]{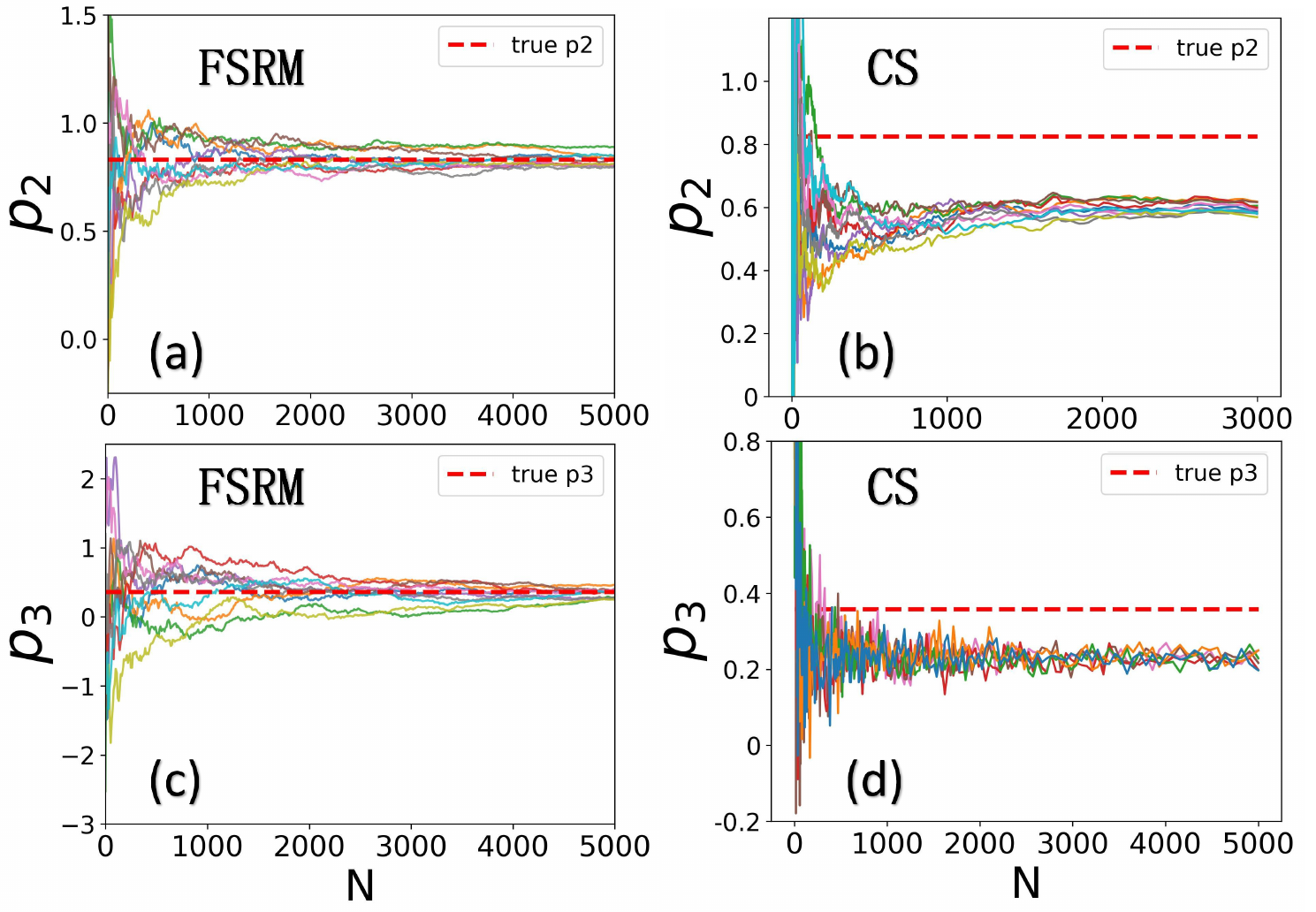}
    \caption{
    \textit{Experimental results of a mixed state by comparing the prediction accuracy for $p_2$ and $p_3$ between the FSRM scheme ((a) and (c)) and CS scheme ((b) and (d)).} 
    For FSRM, we estimate $p_2$ and $p_3$ using 2 and 3 shots per unitary, respectively. Error mitigation technique is only applied to enhance the BM in the estimation of $p_3$. Each experiment is repeated independently ten times, denoted by different lines. For CS, only one shot per unitary is required for the prediction. 
    By comparing the results of (a) and (b), (c) and (d), although both schemes utilize the same dataset, FSRM demonstrates the convergence to the true values while the CS schemes are biased in practice.
    }
    \label{fig:data}
\end{figure}

In the experiment, we collect 3 shots for each measurement setting and unitary. For photon pairs through the outer path, we select 2 shots to calculate $O_2(\mathbf{b}_1,\mathbf{b}_2)$ as the estimation of $p_2$. By repeating this process numerous times, we accurately estimate $p_2$. The results for $p_2$ are depicted in Fig.~\ref{fig:data}(a). Additionally, we select 1 shot from the shots using CM per unitary and combine the unitary information to calculate the classical shadow of the state, $\hat{\rho}=\bigotimes_j (3U^{(j)}|\mathbf{b}^{(j)}\rangle\langle \mathbf{b}^{(j)}|-\id_2)$\cite{huang2020predicting}, which is used for estimating $\hat{p}_2=\mathbb{E}[\tr(\mathbb{S}\hat{\rho}_{i_1}\otimes \hat{\rho}_{i_2})]$ and $\hat{p}_3=\mathbb{E}[\tr(W_{\rightarrow}^A\otimes W_{\leftarrow}^B \hat{\rho}_{i_1}\otimes\hat{\rho}_{i_2}\otimes\hat{\rho}_{i_3})]$ with the CS scheme. Results for CS are presented in Fig.~\ref{fig:data}(b) and (d).

By employing 3 shots in both the outer and inner paths, we estimate $p_3$ with the FSRM scheme by calculating $O_3(\mathbf{b}_1,\mathbf{b}_2,\mathbf{b}_3|\vec{s})$ in Eq.~\eqref{p3_RM}. To mitigate measurement imperfections of BM, we utilize the information from measurement tomography \cite{luis1999complete} to update the function $O_3(\mathbf{b}_1,\mathbf{b}_2,\mathbf{b}_3|\vec{s})$ in Eq.\eqref{Eq_FSRM} by replacing the corresponding Bell basis of $\mathbf{b}_i$ with the real measurement $M_i$. Despite the practical fidelity of BM being less than 80\%, we can make the updated inaccuracy negligible being around $10^{-8}$, and proceed to estimate the true value of $p_3$. Further details on error amendment can be found in Sec. V of Ref.~\cite{supp}.

Results from 10 independent experiments for a mixed state are depicted in Fig.~\ref{fig:data}. The tomography results indicate 2nd PT moment as $p_2=0.858$ and its 3rd one as $p_3=0.3580$. Utilizing the FSRM scheme and predicting from $N_U=10^4$ response numbers, we obtain experimental estimations of $\hat{p}_2 = 0.826 \pm 0.009$ and $\hat{p}_3 = 0.344 \pm 0.034$ (see Fig.~\ref{fig:data}(a) and (c)), converging to the true state values. Additionally, we estimate the mixed state entanglement as $\mc{N}_3=p_2^2-p_3 = 0.338\pm 0.019$ \cite{elben2020mixedstate}. The obtained entanglement value exceeds 0, affirming the existence of mixed-state entanglement. 
However, employing (one-shot) CS scheme using the same dataset as the FSRM scheme, we estimate $\hat{p}_2 = 0.593\pm 0.008$ and $\hat{p}_3=0.229\pm0.010$ (see Fig.~\ref{fig:data}(b) and (d)). 
The results in Fig.~\ref{fig:data}(a) and (c) validates our FSRM methods. And, the data in Fig.~\ref{fig:data}(a) and (b) comes from the same data without any mitigation, and the only difference is CS method ((b)) utilizes the information of unitary while the FSRM method ((a)) does not. 
Our analysis shows that the derivation of predictions of the CS scheme may come from the need for the exact information of the applied random unitaries and thus the unitary errors can affect much; whereas for the FSRM scheme, such kind of information is unnecessary. This discrepancy between CS and FSRM schemes aligns with our analysis regarding Gaussian noise in unitaries, as detailed in Sec.VI of Ref.~\cite{supp}. Consequently, FSRM proves to be more robust for noisy and practical devices.

\begin{figure}[htbp]
    \centering
    \includegraphics[width=0.35\textwidth]{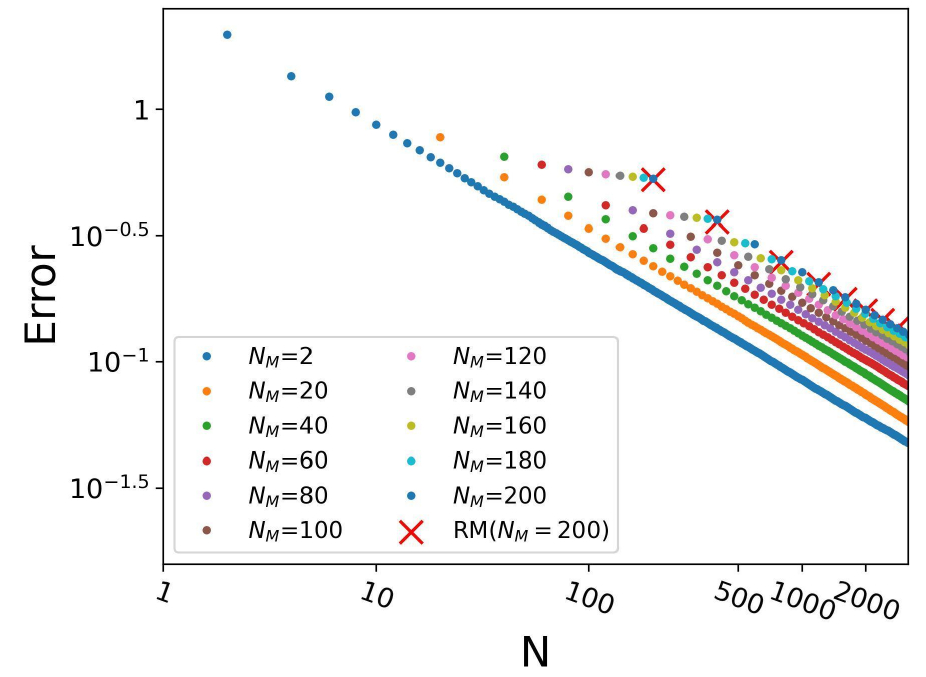}
    \caption{
    \textit{Experimental results of the statistic error performance in prediction of $p_2$ with FSRM.} $N$ for the total measurement time, i.e. the number of photon-pair consumed, and $N_M$ for the shot number, the number of photon-pairs per unitary.
    }
    \label{fig:converging}
\end{figure}

Finally, we investigate the statistical error of the FSRM scheme by taking $p_2$ of a two-photon state as an example. The total measurement time is $N=N_U*N_M$, i.e., the multiplication of rounds and shot-number. The impact of $N_M$ and $N$ on the convergence of the statistical error is illustrated in Fig.~\ref{fig:converging}, in which each line represents the results of over 100 independent experiments. For FSRM, one can get an estimation of every 2-shot from $N_M$-shot under the same unitary, and finally, the result is averaged for all $N_U$ unitaries. For RM, we utilize all $N_M$-shot under the same unitary to compute the conditional probability $\pr(\mb{b}|U)$, which becomes equivalent to our FSRM scheme as $N_M$ tends to be large enough, as shown in Fig.~\ref{fig:converging} for $N_M=200$. Notably, if $N$ is fixed, a smaller $N_M$, i.e., more diversifying unitaries, yields better performance. Through linear regression, we ascertain $\log \mathrm{Err} = -0.5\log N + 0.4393$ as $N_M=2$, and for RM ($N_M=200$), $\log \mathrm{Err} = -0.5\log N +0.8726$, which is labeled by a cross mark in Fig.~\ref{fig:converging}. Consequently, utilizing FSRM results in a 4.3 dB improvement in estimation compared to the original RM. 

\textit{Discussion---}
In this work, we advance the RM framework by integrating BM and few-shot post-processing strategies. We demonstrate its effectiveness in estimating PT moments for challenging mixed-state entanglement quantification.
The BM-enhanced FSRM exhibits robustness to imperfections and noise in random unitary realization, while its few-shot feature ensures efficient post-processing. 
It is worth mentioning that for an n-qubit state, only BM among two qubits is necessary, ensuring scalability for large-scale quantum systems. 
Additionally, we find that, under a fixed measurement resource, the way with more versatile measurement settings with fewer repeats per setting is more efficient, challenging previous estimation protocol with fixed number measurement settings and many repeats per setting.

We expect BM-enhanced FSRM can be further applied to other important scenarios in quantum information and physics, like 
quantum algorithm design \cite{Lubasch2020nonlinear,Yamamoto2021metrology,zhou2022hybrid} and quantum chaos diagnosis \cite{Vermersch2019Scrambling,garcia2021quantum}.
Our research significantly enriches the scope of RM, and advances it in real-world quantum experiments, especially in the characterization of photonic systems. 

\textit{Acknowledgements---}
This work was supported by the Innovation Program for Quantum Science and Technology (Nos. 2021ZD0301200, 2021ZD0301400, 2021ZD0302000), National Natural Science Foundation of China (Grant Nos. 12122410, 12350006, 12205048, 11821404),  Anhui Initiative in Quantum Information Technologies (AHY060300), the Fundamental Research Funds for the Central Universities (Grant No. WK2030000038, WK2470000034).

\bibliography{references}

\end{document}